\documentclass[conference]{IEEEtran} 
\IEEEoverridecommandlockouts
\pdfoutput=1
\usepackage[english]{babel}
\usepackage{graphicx}
\usepackage[latin1]{inputenc}
\usepackage[caption=false]{subfig}
\usepackage{amsmath,amssymb,amsthm,latexsym}
\usepackage{algorithm}
\usepackage{algpseudocode}
\usepackage{xspace}
\usepackage[svgnames]{xcolor}
\usepackage[bookmarks=false,draft=true]{hyperref}
\usepackage{pdfpages}
\usepackage{cite}

\def\MM{\mathcal{M}}

\def\matdesc#1(#2){
\left[\begin{array}{c}#1\end{array}\right]#2
}

\newcommand{\fsize}{\ensuremath{\MM}}
\newcommand{\matdim}[2]{($#1$ x $#2$)}

\newcommand{\seq}[3]{ \ensuremath{{#1}_{#2}, \ldots, {#1}_{#3} }}

\newcommand{\Gmsg}{\ensuremath{\Psi}}

\newcommand{\Gnode}{\ensuremath{\Phi}}

\hypersetup{%
pdftitle={CROSS-MBCR: Exact Minimum Bandwith Coordinated Regenerating Codes},
pdfauthor={Steve Jiekak and Nicolas Le Scouarnec},
pdfkeywords={network coding, regenerating codes, exact repair, multiple failures, coordinated, MBCR},
pdfstartview={FitH},
urlcolor=darkblue,
pdfborder=0 0 0,
bookmarksnumbered
}%

\begin{document}
\sloppy

\definecolor{darkgreen}{rgb}{0,0.5,0}
\def\draft#1{\textcolor{darkgreen}{#1}}


\title{CROSS-MBCR : Exact Minimum Bandwidth Coordinated Regenerating Codes}

\author{
\IEEEauthorblockN{Steve Jiekak}
\IEEEauthorblockA{Technicolor\\
Rennes, France\\
steve.jiekak@epfl.ch}
\and
\IEEEauthorblockN{Nicolas Le Scouarnec}
\IEEEauthorblockA{Technicolor\\
Rennes, France\\
nicolas.lescouarnec@technicolor.com}}
\maketitle

\begin{abstract}
We study the exact and optimal repair of multiple failures in codes for distributed storage. More particularly, we provide an explicit construction of exact minimum bandwidth coordinated regenerating codes (MBCR) for $n=d+t,k,d\ge{}k,t\ge{}1$. Our construction differs from existing constructions by allowing both $t >  1$ (\emph{i.e.}, repair of multiple failures) and $d > k$ (\emph{i.e.}, contacting more than $k$ devices during repair). 
\end{abstract}

\section{Introduction} 
Codes are useful for tolerating failures in distributed storage systems. Yet, erasure correcting codes suffer from huge repair costs after failures. Recently, regenerating codes~\cite{Dimakis2010} have relied on network coding to achieve the optimal tradeoff between storage cost and bandwidth (repair cost). Such codes have latter been extended to support multiple failures (i.e., coordinated or cooperative regenerating codes)~\cite{NetCod2011,Hu2010,Shum2011}. These studies identify two main types of regenerating codes: \emph{(i)} Minimum Storage (MSR/MSCR) regenerating codes which minimize storage costs in priority and \emph{(ii)} Minimum Bandwidth (MBR/MBCR) which minimize bandiwdth costs in priority. Codes achieving these optimal tradeoffs can be built using random linear network codes. Yet, such non-deterministic schemes are not desirable for they require complex integrity checking scheme, and cannot be turned into systematic codes.

Hence, regenerating codes have been studied with the additional constraint of repairing exactly what is lost, in order to build deterministic coding schemes.The problem of repairing exactly a single failure has been well studied both at the MBR point\cite{Rashmi2011} and MSR~\cite{Shah2010,Shah2010b,Rashmi2011,Suh2011,Shah2012} point.  However,  the exact repair of multiple failures remains an open question since it has been studied only  for the very specific setting $d=k$ at both the MSCR point~\cite{Shum2011} and the MBCR point~\cite{Shum2011b}. 

We focus on this problem and propose an explicit code construction for the case of $n=d+t,k,d \ge k,t \ge 1$ thus relaxing the constraint $d=k$ present in previous constructions~\cite{Shum2011b}. In the following, we will present our code construction, the repair algorithm and the decoding algorithm.

\section{CROSS-MBCR Codes}
Minimum Bandwidth Coordinated Regenerating Codes~\cite{NetCod2011} guarantee that a file of $\MM$ original blocks encoded to $n\alpha$ blocks stored uniformly on $n$ devices can be recovered from any $k$ devices, and that if $t$ devices fail the optimal repair procedure consist in downloading $\beta$ blocks from $d$ non-failed devices and $\beta'$ blocks from the $t-1$ other devices being repaired.  If we set $\beta'=1$,\vspace{-1eX}
\begin{align}%
\alpha=2d+t-1 && \beta=2 && \MM=k(2d-k+t)\notag
\end{align}

CROSS-MBCR codes are built upon two encoding matrices:
\begin{itemize}
\item $\Phi$ , which is the generator matrix of $(n,k)$ MDS code (\emph{e.g.}, Cauchy or Vandermonde matrices),
\item $\Psi$ , which is the generator of an $(n-1,d)$  MDS code, such that $(\mathbf{I}_{d,d},\Psi)$ is also the generator of a systematic MDS code (\emph{e.g.}, Vandermonde or Cauchy matrices). 
\end{itemize}
They support any $k$, any $d\ge{}k$, any $t\ge{}1$ and only require that $n=d+t$ (\emph{i.e.}, all devices participate to the repair either as a device providing data or as a device being repaired). In the rest of this section, we will describe the encoding procedure, the repair procedure and the decoding procedure, thus defining completely the code and showing that it satisfy all needed properties.

\begin{figure*}[t]%

\subfloat[Encode]{
	\centering
	\includegraphics[height=0.27\linewidth, trim= 0 0 0 0]{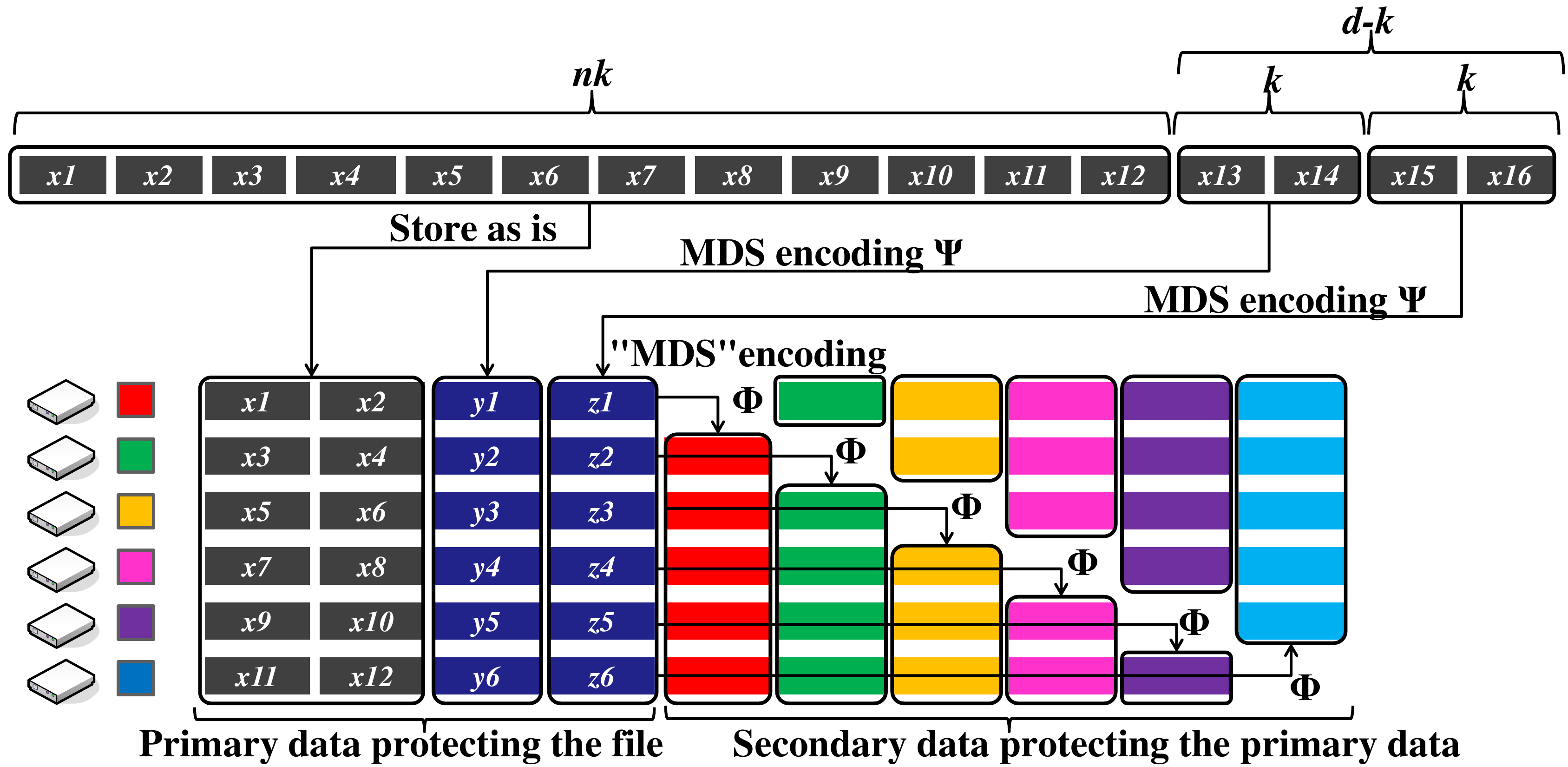}%
		\label{fig:encoding}%
	}\hfill
\subfloat[Decode]{
	\centering
	\includegraphics[height=0.27\linewidth, trim= 0 0 0 0]{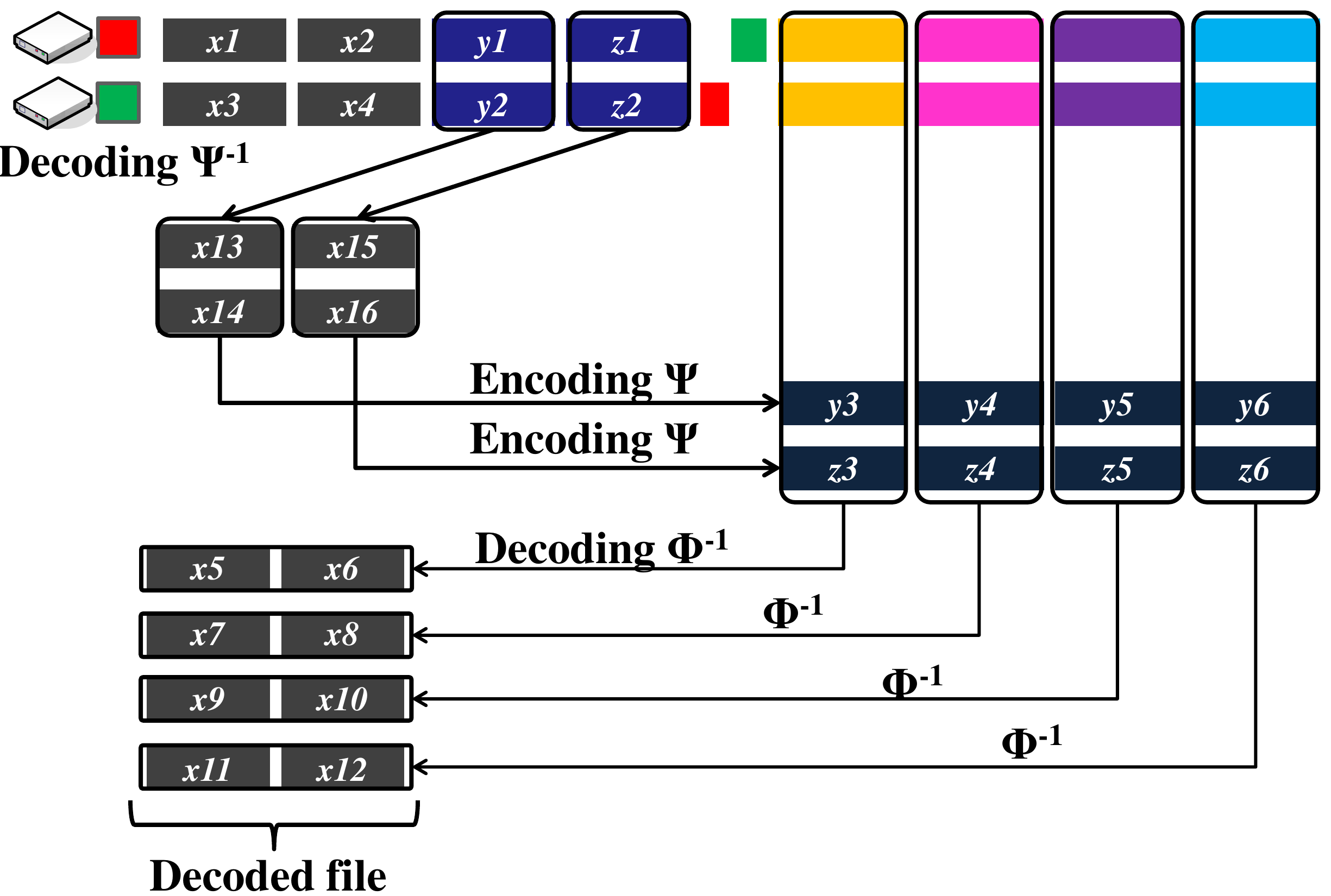}%
		\label{fig:decoding}%
}

\caption{Code for a file stored to $n=6$ devices so that any $k=2$ devices can recover the file, and $t=2$ devices can repair from $d=4$ devices}%
	
\end{figure*}%
\subsection{Encoding}
The encoding procedure is shown on Figure~\ref{fig:encoding}. Given the condition $n=d+t$, the file size \fsize{} can be rewritten as $kn+k(d-k)$. The file $(\seq{x}{1}{\fsize})$ is divided in $n$ sequences $\seq{a}{1}{n}$ and $d - k$ sequences $\seq{b}{1}{d - k}$ such that\vspace{-2eX}
\begin{align*}
 a_{i} &= (\seq{x}{(i-1)k+1}{ik})_{i=1\ldots n}\\
 b_{j} &= (\seq{x}{(j-1)k+ 1 + kn }{jk + kn})_{j=1\ldots d - k}
\end{align*}

\paragraph*{Step 1}
The sequence $a_i$ is  written on the first $k$ positions of the device $i$. Each sequence $b_j$ is encoded using the generator $\Gmsg$, and the $i^\mathrm{th}$ resulting block is stored at the $k+j$ position of device $i$. Let us name $w_i=(a_i\;\Gmsg_i^tB)$ the sequence of blocks stored on the $d$ first positions of device $i$, where  $B$ is the \matdim{k}{d-k} matrix with $b_j$ as column vectors. These $d$ blocks are designated as primary data blocks.


\paragraph*{Step 2}
The primary data blocks $w_i$ stored on device $i$ are encoded using $\Gnode$ and the resulting $n-1$ blocks are stored on all other devices.
Each device stores a total of $n-1$  blocks $    p_i= \begin{pmatrix} 
	        \Gnode{}_{1}^{t}w_{i\oplus 1} & 
	        \Gnode{}_{2}^{t}w_{i\oplus 2} & 
	        \ldots & 
	        \Gnode{}_{n-1}^{t}w_{i\oplus n-1} 
	      \end{pmatrix}$ where $\oplus$ is the addition modulo $n$. These blocks are designated as secondary data blocks. The technique used for building these secondary block is similar to the one used in~\cite{Shum2011b}.

\subsection{Repairing}
\begin{figure}[h]%
\centering
	\includegraphics[width=\linewidth, trim= 0 0 0 0]{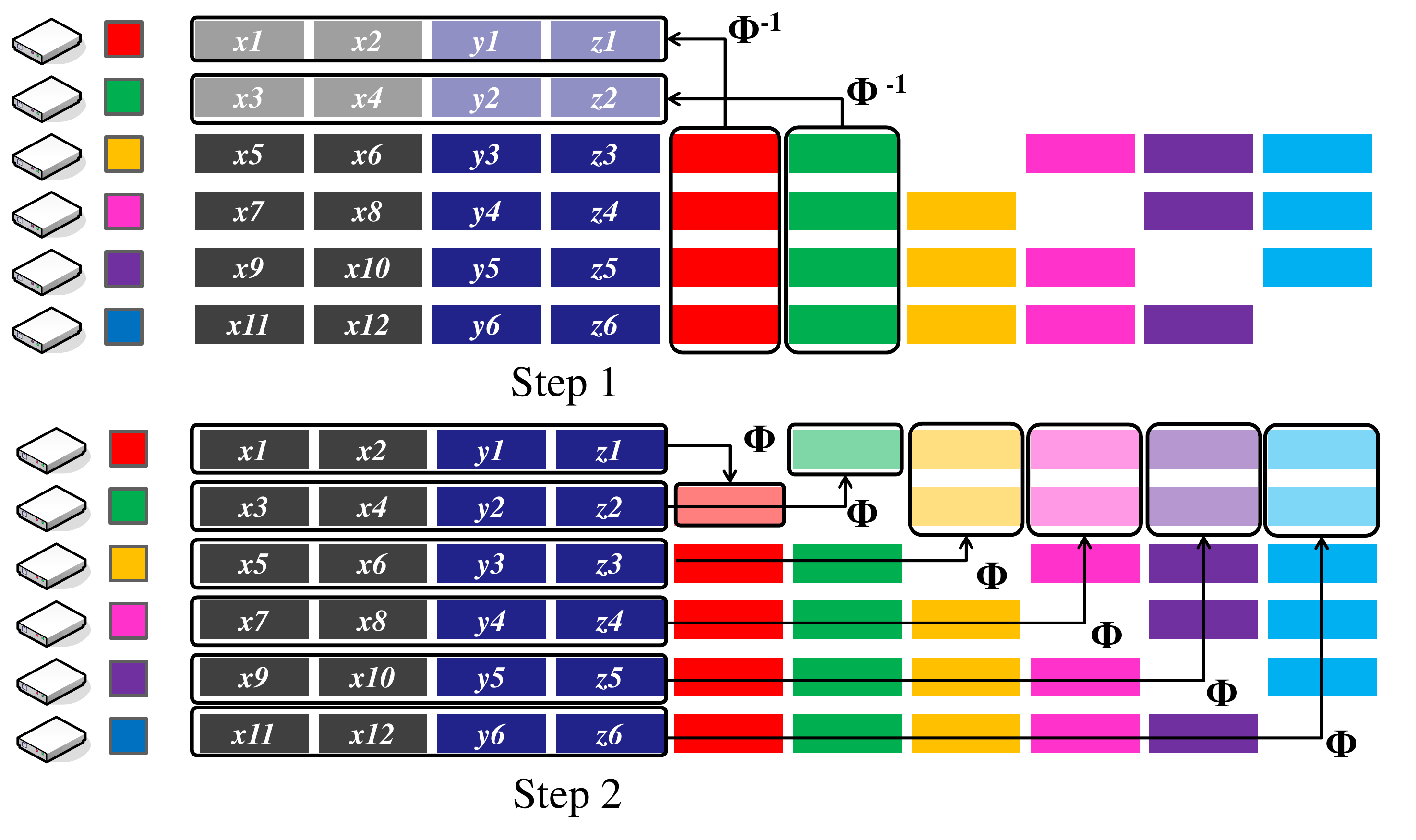}%
\caption{Repairing a code $(n=6, k=2, d=4, t=2)$}%
		\label{fig:repairing}%
\end{figure}

The repair procedure is illustrated on Figure~\ref{fig:repairing}.

\paragraph*{Step 1} 
Each  device being repaired $i$ fetches the secondary data blocks $\Gnode{}w_i$ still available from live devices. Since $\Gnode$ is the encoding matrix of an MDS code, the device being repaired can decode and recover the primary data blocks $w_i$. 

\paragraph*{Step 2} The missing secondary data blocks, initially stored on failed devices, are re-generated from primary data blocks as done in the second step of the encoding.

\subsection{Decoding}
The decoding procedure is shown on Figure~\ref{fig:decoding} and use both the primary data and the secondary data.


\paragraph*{Step 1} The primary data is retrieved from the $k$ contacted devices. The encoded part is decoded using $\Psi^{-1}$ since $\Psi{}$ is the generator matrix of an $(n,k)$ MDS code. A total of $k(k + d-k)$ original blocks are recovered during this step. 

\paragraph*{Step 2} The part decoded during step 1 is encoded again using  $\Psi$ to recreate the encoded part of the primary data blocks stored on the $n-k$ non contacted devices.

\paragraph*{Step 3} The secondary data blocks retrieved from the $k$ contacted devices and the encoded primary data blocks recreated at step 2 are processed together and decoded using $\Phi^{-1}$ to recover the remaining $k(n-k)$ original blocks. This is possible since $(I_{d,d}\; \Phi{})$ is the generator matrix of an $(n-1+d,d)$ systematic MDS code.  As a consequence, the file can be fully recovered from any $k$ devices.

\section{Conclusion}
In this paper, we define an explicit exact MBCR codes construction, thus showing that it is possible to build exact minimum bandwidth coordinated regenerating codes for a wide set of parameters ($n=d+t,k,d,t$). It is interesting to notice that if $d=k$, the sequences $b_j$ encoded using $\Psi$ disappear, and the whole scheme degenerates into the same coding scheme as initially proposed for $d=k$ by Shum et al. \cite{Shum2011}. As a consequence, our scheme encompass the single previous exact MBCR codes construction as a special case.

\vfill
\bibliographystyle{IEEEtran}
\bibliography{regenerating}

\includepdf[frame=false,fitpaper=false]{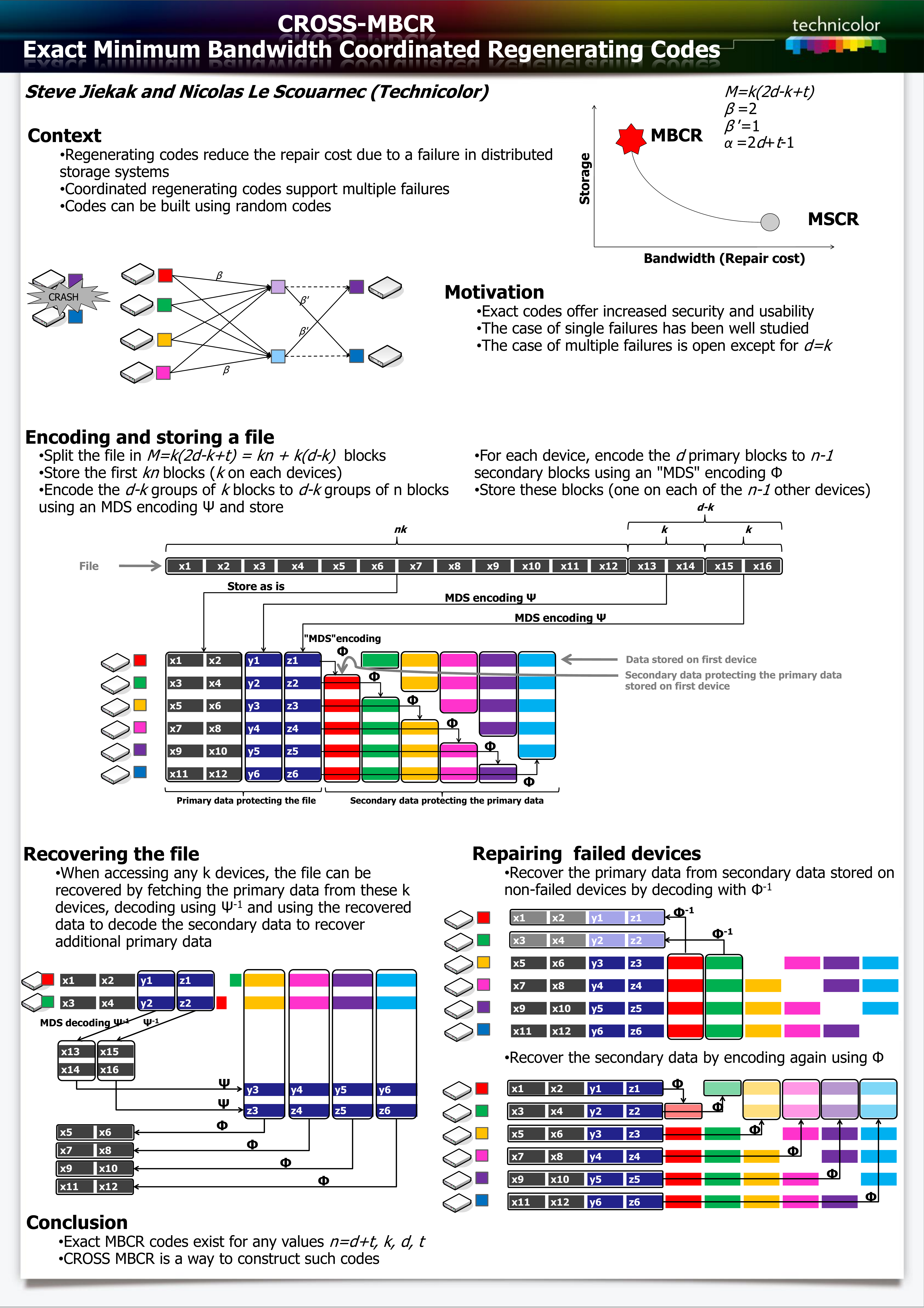}
\end{document}